\documentclass[%
aip,
amsmath,amssymb,
reprint,%
]{revtex4-1}

\usepackage{graphicx}% Include figure files
\usepackage{dcolumn}% Align table columns on decimal point
\usepackage{bm}% bold math
\usepackage[utf8]{inputenc}
\usepackage[T1]{fontenc}
\usepackage{mathptmx}
\usepackage{lmodern}

\newcommand{\affA}{CNRS, Aix-Marseille Universit\'{e}, IM2NP (UMR 7334),
Institut Mat\'{e}riaux Micro\'{e}lectronique et Nanosciences de Provence, 
Marseille, France.}
\newcommand{\affB}{Department of Chemistry, Florida State University,
Tallahassee, Florida 32310, USA.}
\newcommand{\affC}{CNRS, Aix-Marseille Universit\'{e},  Centrale Marseille, ISM2, Institut des science mol\'{e}culaire de marseille,  Marseille, France.}
\newcommand{\affD}{Aix Marseille  Universit\'{e}, CNRS, Centrale Marseille, FSCM, Spectropole, Marseille, France}
\newcommand{\affE}{The National High Magnetic Field Laboratory, Tallahassee,Florida 32310, USA.}

\newcommand{\Mg}{DMAMgF:Mn$^{2+}$}
\newcommand{\Zn}{DMAZnF:Mn$^{2+}$}
\begin{document}

\preprint{AIP/123-QED}

\title{Quantum dynamics of Mn$^{2+}$ in dimethylammonium magnesium formate }
% Force line breaks with \\

\author{M. Orio}  \email{maylis.orio@univ-amu.fr.}\affiliation{ \affC}%
\author{J. K. Bindra}\affiliation{ \affB}\affiliation{ \affE}%
\author{J. vanTol}\affiliation{ \affE}%

\author{M. Giorgi}\affiliation{\affD}
\author{N. Dalal}\affiliation{ \affB}\affiliation{ \affE}%

\author{S. Bertaina}  \email{sylvain.bertaina@im2np.fr.} \affiliation{\affA}

\date{\today}% It is always \today, today,
%  but any date may be explicitly specified

\begin{abstract}
	Dimethylammonium magnesium formate,  [(CH$_3$)$_2$NH$_2$][Mg(HCOO)$_3$] or DMAMgF, is a model to study high temperature hybrid perovskite-like dielectrics. This compound displays a phase transition from para to ferroelectric at about 260~K. Using multifrequency electron spin resonance in continuous wave and pulsed modes, we herein present the quantum dynamic of Mn$^{2+}$ ion probe in DMAMgF. In the high temperature paraelectric phase, we observe a large distribution of the zero field splitting that is attributed to high local disorder and further supported by DFT computations. In the low temperature ferroelectric phase, a single structure phase is detected and shown to contain two magnetic structures. The complex EPR signals were identifed by the means of Rabi oscillation method combined to crystal fields kernel density estimation.  
\end{abstract}

\maketitle

\section{Introduction}

\begin{figure*}
	\centering
	\includegraphics[width=1\linewidth]{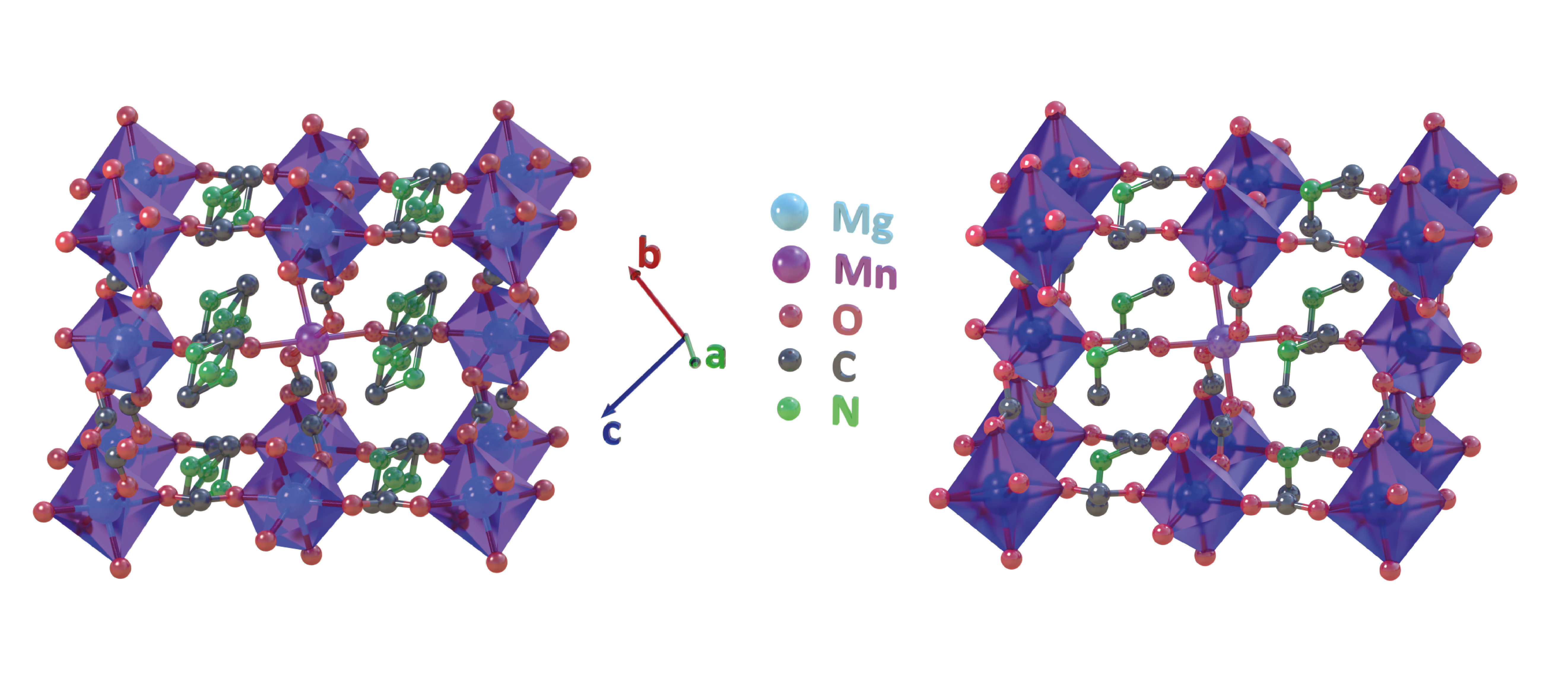}
	\caption{Crystal structure of the DMAMgF framework at $T > T_c$ (left, HT
		phase - R$\bar{3}c$) and $T  < T_c$ (right, LT phase - $Cc$). The Mg (in
		blue)
		and Mn probe (in pink) are in the center of the oxygen octahedron. The DMA$^+$
		are located in the formate cages and have three equivalent positions in the HT phase,
		represented by 3 nitrogens (green) while in the LT phase only one position is
		expected. For clarity the hydrogen atoms are not shown. The figures are
		realized
		with Blender \footnote{3D open source software : https://www.blender.org/}
		using the structure investigated in this work. }
	\label{fig:structure}
\end{figure*}

	Metal-organic frameworks (MOFs) are constituted of two main building units: the
framework consisting of metal centers connected to each other by organic linker
molecules and the cation molecule in the cavity the dynamics of which are
responsible for the dielectric transition. Previous works report hybrid
organic-inorganic metal-organic frameworks (MOF), such as
[(CH$_3$)$_2$NH$_2$][M(HCOO)$_3$] (dimethylammonium metal formate or DMAMF, M
is divalent transition metal
ions\cite{Jain2008,Besara2011,Jain2009,Abhyankar2015,Bertaina2018,Wang2013,Sanchez-Andujar2014}. Heat capacity and dielectric measurements of DMZnF
indicated a phase transition at approximately 160~K \cite{Besara2011}. The
order-disorder phase transition is a common property in these materials. At
higher temperatures in the disordered phase the (CH$_3$)$_2$NH$_2^+$ (DMA$^+$,
dimethylammonium) cation that is trapped within the cage is disordered, which
means that the nitrogen from the amine group can
occupy three locally equivalent positions by forming hydrogen bonds with oxygen
atoms from the formate linkers. As the temperature is decreased, the long-range
order is established due to the cooperative ordering of the cations at $T_c$.
In the low temperature phase the nitrogen atoms freezes in to a single position
in the cavity, while the metal-formate framework becomes more distorted
\cite{Besara2011}. The Mg analogue (DMMgF), is known to exhibit dielectric
transition at exceptionally high $T_c$ of approximately 270
K\cite{Pato-Doldan2012}. The single crystal X-ray diffraction (XRD) studies reveal that the
high- and low-temperature
phases of these compounds belong to the trigonal, $R\bar{3}c$ (centrosymmetric)
and the monoclinic, $Cc$ (non-centrosymmetric) space groups, respectively
\cite{Pato-Doldan2012,Asaji2014}. The metal-formate frameworks of these
materials consist of pseudo-cuboid cavities, each containing a single DMA+
cation as shwo in Fig. \ref{fig:structure}. DMMgF has been intensively
investigated using heat capacity, dielectric, and XRD
measurements\cite{Pato-Doldan2012,Asaji2014}. However, despite this huge effort
the precise nature of the phase transition in DMMgF is still obscured. Among
many other experimental methods, electron paramagnetic resonance (EPR)
spectroscopy is well-suited to study structural phase transitions
\cite{Abhyankar2015,Abhyankar2018,Bertaina2018,Simenas2015,Simenas2018,Simenas2017}.
It used to detect the local environment of a paramagnetic center (e.g., local
order parameter such as electric polarization) that can be influenced by the
structural transformations. Although most of the MOFs do not contain any
intrinsic paramagnetic center, they can be doped with a small amount of
paramagnetic transition metal ions (e.g., Mn$^{2+}$) which act as local probes in the
structure. In our previous studies we have employed continuous wave (CW) EPR
spectroscopy to successfully investigate the phase transition in undoped DMAMnF 
and \Zn MOFs \cite{Abhyankar2015,Bertaina2018}. However, due to the strong
magnetic dipolar and exchange interactions between the Mn$^{2+}$ centers the EPR
spectrum of DMAMnF consists of a single broad line which is barely sensitive to
the ordering of the DMA$^+$ cations \cite{Abhyankar2015}. Also, slow dynamics of
DMA$^+$ cation around the phase transitions in \Zn was investigated using S-Band
(4~GHz) EPR. S-Band EPR spectra yielded clear signatures of the slow motion of
both the formate and DMA$^+$ groups\cite{Bertaina2018}. EPR methods have also
been used to characterize the low temperature phases and dynamics in Mn$^{2+}$ and
Cu$^{2+}$ doped niccolite [NH$_3$(CH$_2$)$_4$NH$_3$][Zn(HCOO)$_3$]$_2$
\cite{Simenas2018} . Mn$^{2+}$ doping in such systems is the probe of choice for
the  local properties (crystal fields, motion). The large number of EPR
transitions as well as the long coherence time allows us to have access to the
quantum dynamics of the spins. Large spin ions like Mn$^{2+}$ ($s=5/2$) is used
for their quantum coherence properties as a potential
qubit\cite{Loss1998,Leuenberger2001}. In weak crystal field it was shown that
multiple quantum coherence can be induced and controlled \cite{Bertaina2009,Bertaina2015a, Bertaina2011}. In this paper, by mean of incoherent (continous
wave) and coherent (pulsed wave) EPR, we describe the quantum dynamics of \Mg.
We show how the complex quantum dynamics of Mn$^{2+}$ in moderate crystal fields
can be resolved using a kernel density estimation.

\section{Materials and methods}
\subsection{Sample preparation}
 Sample preparation The DMAMgF:Mn2+samples were prepared using the method
similar to the one described in detail in our earlier paper with some minor
modifications. A 23 mL solution of 50 vol \% dimethylformamide (DMF) in Nanopure
water into which 85.0 mM MgCl$_2$ and 2 $\mu$M MnCl$_2$ (0.002\%) were dissolved was
sealed in a 35~mL pressure vessel. The pressure vessel was then heated to
140~$^circ$C for 2 days, then allowed to cool to room temperature. Once cool,
the supernatant was decanted. Single crystals were obtained by slowly
evaporating the solution in a 10 dram glass vial with a 1 in. diameter. We
assumed a cuboid shape of the crystal, and the axes of the Cartesian reference
frame were chosen such that they coincide with the edges of the crystal. (102)
parallel to H and the angular dependent single crystal HF-EPR measurements were
performed by rotating the sample about (112) plane.

\subsection{Single crystal X-ray diffraction}
A suitable crystal for compound \Mg was measured on a Rigaku Oxford Diffraction
SuperNova diffractometer at 220 K ($<T_c$=260K) at the CuK$\alpha$ radiation
($\lambda$=1.54184 \AA). Data collection reduction and multiscan ABSPACK
correction were performed with CrysAlisPro (Rigaku Oxford Diffraction). Using
Olex2\cite{Dolomanov2009} the structures were solved by intrinsic phasing
methods with SHELXT \cite{Sheldrick2015}  and SHELXL\cite{Sheldrick2015a} was
used for full matrix least squares refinement. All H-atoms were found
experimentally and their coordinates and Uiso parameters were constraint to
1.5Ueq (parent atoms) for the methyls and to 1.2Ueq (parent atom) for the other
carbons.

\subsection{EPR spectroscopy}

X-band (about 9.6GHz) EPR measurements were performed using two conventional
Bruker spectrometers operating in continuous wave (cw) mode  X-band (EMX - 9.6
GHz) and pulse mode X-band (E680 9.6 GHz). The cw spectrometer uses a standard
4102ST resonator (TE$_{102}$) installed in an oxford cryostat ESR900. Low
temperature measurements were performed using a cryogen free Bruker Stinger
cryocooler allowing the temperature to reach 7K. The angular dependence was
measured using an automatic goniometer. Magnetic field modulation
($f_m=100$~kHz) associated with lock-in detection was employed, resulting in the
derivative of the signal. The amplitude of the modulation can be set up to 10~G
and was carefully chosen to be below any linewidth to avoid overmodulation
effect. 

The pulse spectrometer was used to performed  Rabi oscillations measurements on
Mn$^{2+}$ ions \cite{Bertaina2011} using the sequence
$P_R-\tau_1-\pi/2-\tau_2-\pi-echo$ where $P_R$ is Rabi pulse which controls the
coherent rotation of the spin and $\tau_1$ is a wait time longer than the coherence
time in order to destroy the transverse magnetization $T_2$ but shorter than the
relaxation time $T_1$ maintaining the longitudinal magnetization $\left\langle S_z\right\rangle $. The later
is then recorded by the standard Hahn echo.  Du to the large distribution of
transitions, the $\pi/2$ pulse is selective in \Mg. However, it is more convenient in
field sweep Rabi oscillation sequence since one want to excite just a small
quantity of spin (the ones actually in resonance) in the spin packet. The
measurements were performed at 7K using the Bruker MD5 dielectric resonator
overcoupled. The microwave field $h_{mw}$ was calibrated by measuring the
nutation of a S=1/2 standard. 

High-field/high-frequency EPR (HF-EPR) experiments have been carried out
using a homemade quasioptical superheterodyn setup developed at NHMFL
\cite{vanTol2005}. The spectrometer operates at  240~GHz and
at temperature from RT down to 5~K. Angular dependence with respect of magnetic
field direction is achieved  using a manual goniometer every 18$^\circ$

\subsection{DFT}
\begin{figure}[t]
	\centering
	\includegraphics[width=\columnwidth]{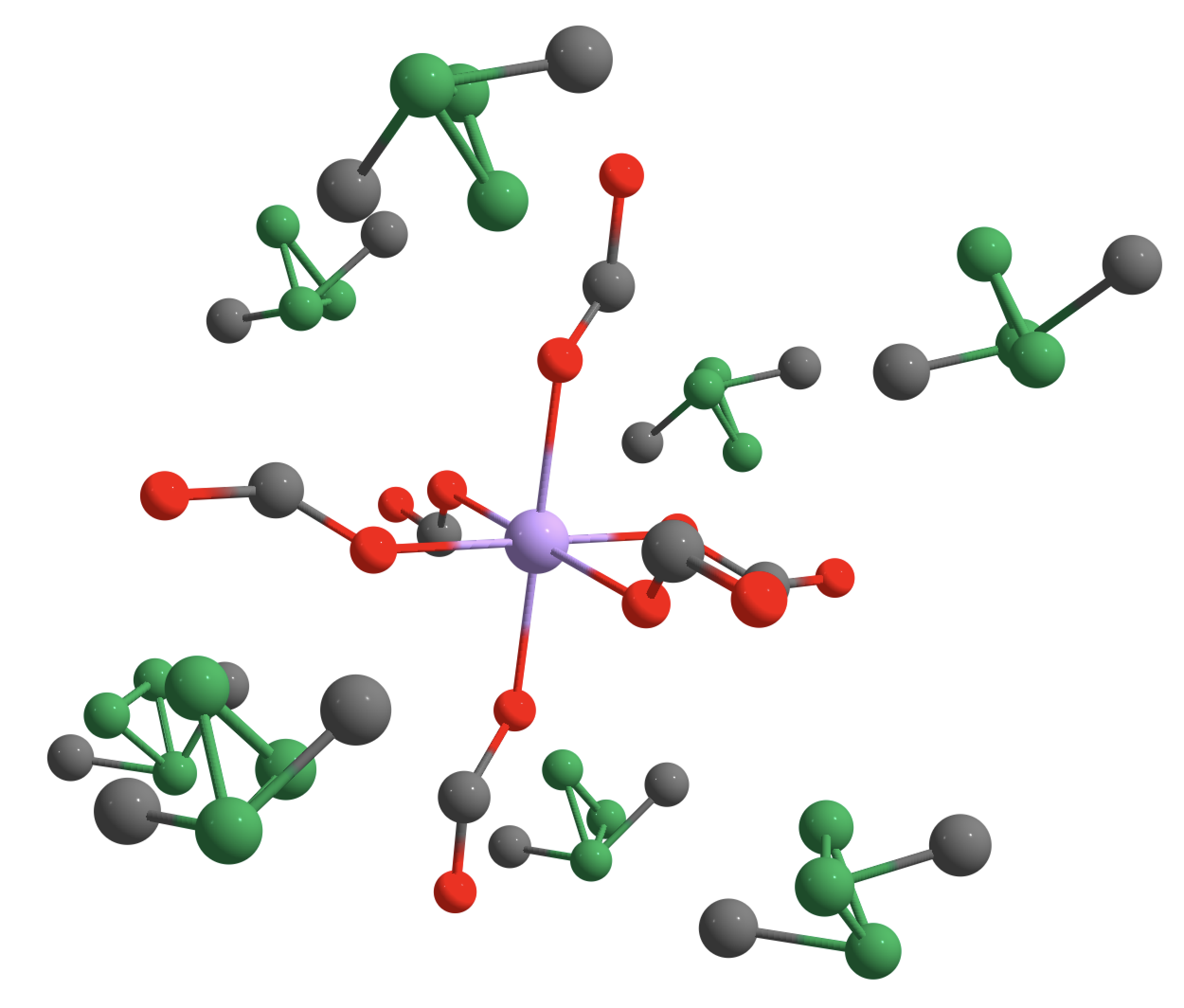}
	\caption{Minimal model used for computing the zfs parameters in DMMgF showing
		equivalent positions of the nitrogens in the DMA+ cations. Color code Mn:
		pink, N: green, O: red, C: dark gray, H: omitted for clarity.}
	\label{fig:DFT}
\end{figure}
All theoretical calculations were based on the Density Functional Theory (DFT)
and were performed with the ORCA program package\cite{Neese2018}. To facilitate
comparisons between theory and experiments, all DFT models were obtained from
the experimental X-ray crystal structures and were optimized while constraining
the positions of all heavy atoms to their experimentally derived coordinates.
Only the positions of the hydrogen atoms were relaxed because these are not
reliably determined from the X-ray structure. Geometry optimizations were
undertaken using the GGA functional BP86\cite{Perdew1986,Perdew1986a,Becke1988}
in combination with the TZV/P\cite{Schafer1994} basis set for all atoms and by
exploiting the resolution of the identity (RI) approximation in the Split-RI-J
variant\cite{Neese2003} with the appropriate Coulomb fitting
sets\cite{Weigend2006}. Increased integration grids (Grid4 and GridX4 in ORCA
convention) and tight SCF convergence criteria were used. The zero-field
splitting parameters were obtained from additional single-point calculations
using the BP functional. Scalar relativistic effects were included with ZORA
paired with the SARC def2-TZVP(-f) basis sets\cite{Pantazis2008,Pantazis2009}
and the decontracted def2-TZVP/J Coulomb fitting basis sets for all atoms. The
spin-spin contribution to the zfs was calculated on the basis of the UNO
determinant\cite{Sinnecker2006}.

\section{Rabi distribution calculation}

To simulate the EPR spectra, we used the following Hamiltonian:

\begin{equation}\label{eq:Ham1}
	\mathcal{H}=\mu_b \vec{H}[g]\vec{S} + \vec{S}[A]\vec{I} + \sum_k\sum_qB_k^q
	\hat{O}_k^q(\vec{S})
\end{equation}
Here, the first term represents the Zeeman interaction with $[g]$ the g tensor
and $\mu_b$ is the Bohr magneton, the second represents the hyperfine
interaction with the hyperfine constant tensor $[A]$ considered isotropic, while
the last term represents the crystal-field interaction in the formalism of the
extended Stevens operators\cite{Abragam1950,Ryabov1999} $\hat{O}^q_k(\vec{S})$
($k$=2, 4, 6 and $q=-k, \dots, k$ ). The $B_k^q$ are real coefficients with the
relations : $3B_2^0=D$ the axial anisotropy, $B^2_2 = E$ the rhombic anisotropy,
$24B^4_4 = a$ the cubic contribution and $F=180B^0_4-36B_4^4$ the fourth order
contribution. $k=6$ terms were considered small enough to be neglected. $[g]$
and $[A]$ are considered as scalar (isotropic) for Mn$^{2+}$ .

The EPR simulations were performed using a hybrid method. The conventional cw
EPR spectra were simulated using the Matlab package Easyspin v5.2.28
\cite{Stoll2006} .

The Rabi mode distributions were computed using a database approach. Due to the
large anisotropy and the disoriented nature of \Mg at low temperature, the full
dynamical density matrix model developed in ref. [\onlinecite{Bertaina2015}] for
$n$photon transitions should have been too heavy to implement and  unnecessary.
Due to the large zfs expected in this family of compounds \cite{Simenas2018,
	Bertaina2018}, only the 1-photon transitions are expected to occur(n-photon transitions
are expected when $D\sim h_{mw}$\cite{Bertaina2011a} with $h_{mw}$ the microwave
field ). Using the first order Fermi golden rule, the Rabi frequency of a
transition $m$ to $n$  is $|<m|S^+|n>|$. The database of the Rabi frequencies
was constructed as the following. The static field orientation is set first,
then, after diagonalization of the Hamiltonian,  all transitions fields are
computed (regardless of the intensity), the Rabi frequency of each transition is
calculated by the Fermi's golden rule and the intensity of the  transition 
simply equals the square of the Rabi frequencies. Orientations, resonance fields,
transitions and Rabi frequencies are collected in the database. The treatment of
the data is then realized using Pandas module of Python 3.8. We use the kernel
density estimation (KDE)\cite{Rosenblatt1956} to reconstruct the Rabi
oscillation distribution:

\begin{equation}\label{eq:KDE}
	\hat{f_h}(x)=\frac{1}{n}\sum_{i=1}^nK_h(x-x_i)
\end{equation}
where $K$ is the kernel function, $h$ the bandwidth and $n$ the number of
samples. This method can be seen as an extension of the histogram method which
counts the number of occurrences  around a value. In our analysis we used
the Gaussian kernel function .
Since we have access to the transitions, field resonances and orientations, we
can identify the nature and distribution of all transitions.

\section{Results and discussion}

\subsection{HT Phase}

\begin{figure}[t]
	\centering
	\includegraphics{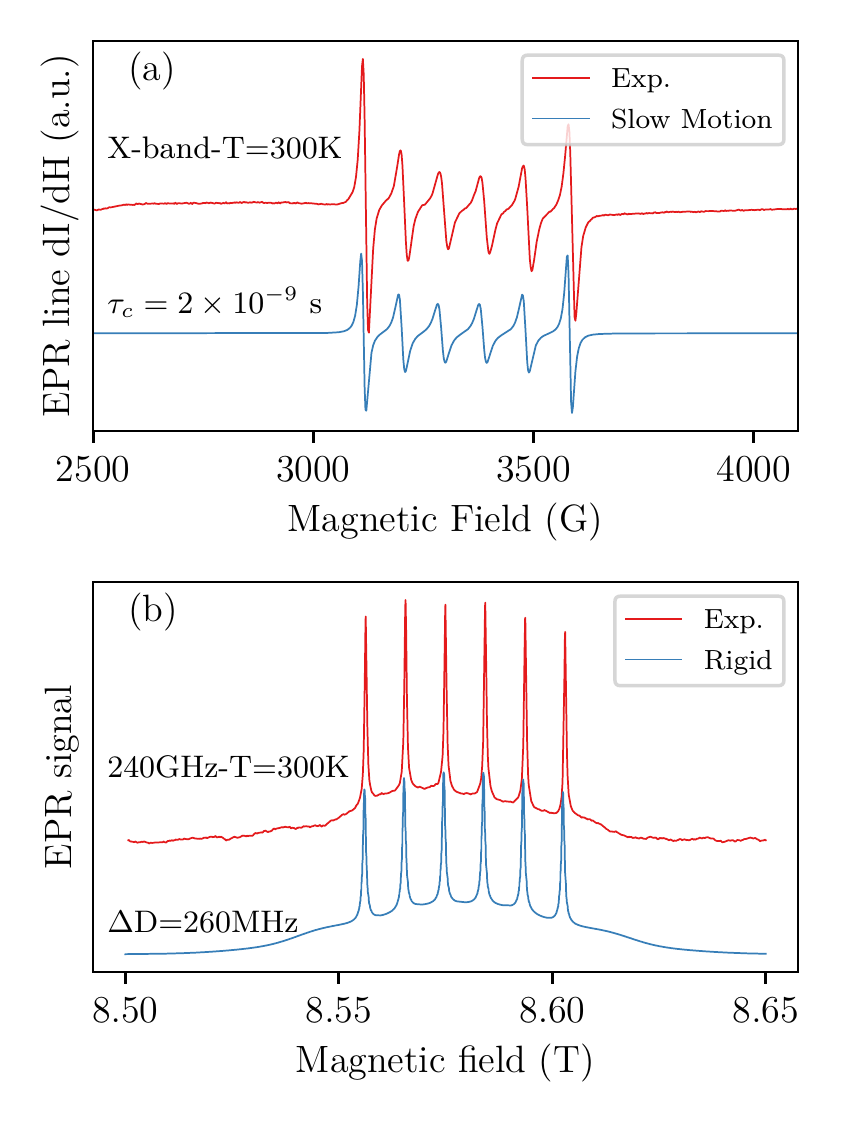}
	\caption{(color online) (a) First derivative with respect to the field of the EPR
signal of a single crystal of \Mg recorded at X-band and room temperature
(red). The blue line below represents the simulation of the spectrum in the
slow motion regime with a correlation time $\tau_c=2\times 10^{-9}$s. (b) EPR
signal recorded on the same crystal at $f_{mw}$=240~GHz and at room
temperature. The blue line is the rigid limit spectrum simulation using a
distribution of axial anisotropy $\Delta D$.}
	\label{fig:eprx240ghz}
\end{figure}

Continuous wave EPR studies have been previously reported for the LT phase of the Zn compound,
DMAZnF:Mn$^{2+}$ \cite{Simenas2015,Simenas2018} and a particular
focus has been made on the dynamics in the HT phase \cite{Bertaina2018}. Here we
are interested in determining focus on how the change from Zn to Mg, which highly affects the structural phase
transition temperature, can also have effects on the dynamics probed by EPR.
Figure \ref{fig:eprx240ghz}(a) shows the EPR signal of \Mg for H$\|$[102] at
room temperature. The signal is composed of 6 lines corresponding to the $m_i$
nuclear spin projection of $^{55}$Mn. The separation between lines is 93.6 G
which corresponds to 262.3 MHz or 87.5$\times 10^{-4}$ cm$^{-1}$ and is a
classical value reported for Mn$^{2+}$ ions \cite{Low1957}. In solid-state EPR
and for single crystal measurements, the intensity and linewidth of the 6 line pattern is expected
to be comparable while in our case the lines at lowest and highest field (corresponding
to $m_i=\pm 5/2$ ) are clearly sharper than the onespresent in the intermediate field region. Moreover no
forbidden transition \cite{Smith1968} is resolved while it is usually observed
in  Mn$^{2+}$ ions in anisotropic crystal field. This behavior was also
observed\cite{Bertaina2018}  in DMAZnF:Mn$^{2+}$ and was attributed to the slow
motion regime \cite{McLachlan1964,Goldammer1974,vonGoldammer1974} caused by
the DMA$^+$ movement around the Mn$^{2+}$ ion. Using this model\cite{McLachlan1964}
we managed to simulate the experimental data within DMAZnF:Mn$^{2+}$. Indeed, by
decreasing the temperature, the DMA$^+$ slow down and when its correlation time
$\tau_c$ is longer than the time scale of the measurement (i.e. $> 1/f_{mw}$), the
system is considered frozen and can be described by the rigid model. However, in
\Mg, $T_c$ (263~K) is higher than in the Zn counterparts (173~K) and the structure of the system changes
before reaching the frozen regime. To observe the frozen regime we
have to increase significantly the frequency. Fig. \ref{fig:eprx240ghz}(b) shows
the EPR signal of \Mg recorded at $f_{mw}$=240~GHz and at room temperature. Contrary to the X-band measurements, all of the six nuclear isotope lines for
transitions $m_s= -1/2 \leftrightarrow 1/2$ have the same intensity suggesting
that we have reached to rigid limit of EPR. However, whatever the orientation of
crystal is, no satellite lines corresponding to $m_s= \pm 5/2 \leftrightarrow
\pm 3/2$  and   $m_s= \pm 3/2 \leftrightarrow \pm 1/2$ were resolved indicating
a large distribution of the crystal field parameters. To simulate the spectrum
of  Fig.\ref{fig:eprx240ghz}(b) we used the crystal field parameters of the \Zn compound
\cite{Bertaina2018} $D=B^2_0/3$=250~MHz but we have to set a distribution of the
crystal field parameter $\Delta D$=260~MHz which seems inconsistent since $\Delta
D$ is usually less than 10\% of $D$. To explain this large value of $\Delta D$ it is worth to mention that,
in the HT phase, the system is locally highly disordered. Indeed, the Mn$^{2+}$ ion is
surrounded by 8 DMA$^+$ cations which all have 3 different positions giving rise to
$3^8=6561$ configurations of the crystal field which are responsible for the large zfs distribution.

To support and rationalise the experimental findings about $\Delta D$, DFT
calculations were conducted. To do so, we employed a methodology similar to that
from our previous study on \Zn \cite{Bertaina2008} and worked with a minimal model
consisting in one Mn$^{2+}$ ion bound to 6 formate anions and surrounded by 8
DMA$^+$ cations \footnote[1]{See Supplemental Material}.  The resulting metal cluster displays a quasi-octahedral
coordination geometry. Based on the high temperature single crystal XRD structure of DMMgF
that identified three equivalent positions of the nitrogen in each DMA+, we have
considered several configurations in which the Mn-N distances for 6 DMA$^+$ can
take values of 4.495, 5.121 and 5.679  \AA{} while the 2 remaining	DMA$^+$ display Mn-N distances of 5.698 \AA{}. This provides a random sampling
of the different situations and allows to determine the distribution of the zfs
parameter, which was found to be $\Delta D \approx 234$~MHz. The computed value
is in fair agreement with the experimentally estimated value of 260~MHz and our
calculations adequately reproduce the increased value for the zfs distribution
when comparing \Mg to \Zn ($\Delta D_{DFT} \approx 125$~MHz and
$\Delta D_{exp} \approx 150$~MHz). While there is no clear evidence for an
effect
from the Mg, our results confirm the influence from the DMA$^+$ cations on the
crystal-field effect on the Mn$^{2+}$ as observed in the case of \Zn.

\begin{figure}[h]
	\centering
	\includegraphics{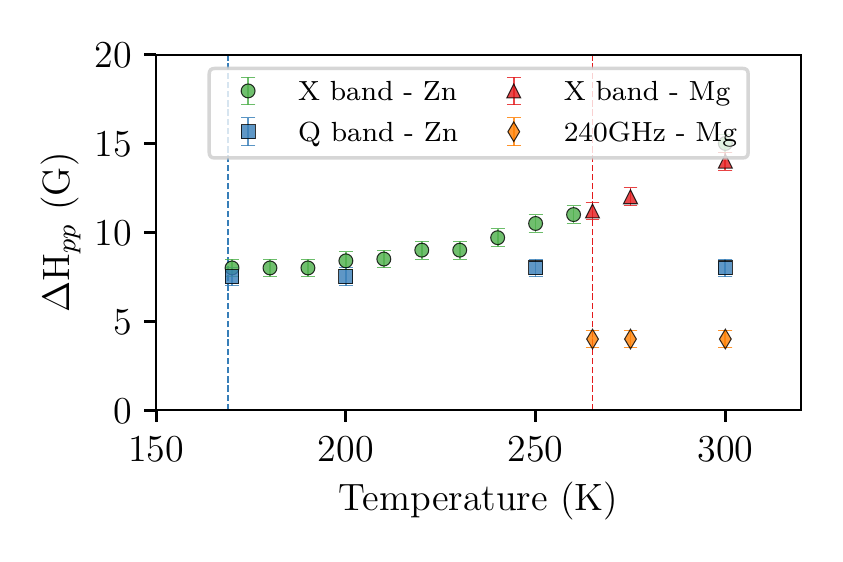}
	\caption{Peak-to-peak linewidth of the rightmost EPR line in \Mg, recorded
		between 300 and 260~K at X-band and 240~GHz. The data for \Zn are extracted
		from
		Ref. [\onlinecite{Bertaina2018}]. The vertical dashed line is the structural
		transition temperature. }
	\label{fig:linewidthtemperature}
\end{figure}

Fig. \ref{fig:linewidthtemperature} shows the temperature dependence of the
peak-peak linewidth $\Delta H_{pp}$ of the smallest field line ($m_I=-5/2$) for
\Mg (this work) and \Zn (from Ref. [\onlinecite{Bertaina2018}]) in the HT phase.
In \Zn, $T_c$=170~K while in \Mg   $T_c$=260~K that's why the temperature range of
the HT phase in \Mg is more limited. In X-band and in slow regime,  the temperature
dependence of $\Delta H_{pp}$ is the barely the same in the two compounds
showing that the correlation time $\tau_c$ of the DMA$^+$ cation is weakly affected by the
nature of the metal, Zn or Mg, as pointed  out in Ref. [\onlinecite{RamakrishnaSanath2021}]. In the rigid limit the linewidth of the $m_s= - 1/2 \leftrightarrow  1/2$ 
transition in \Mg is about two times smaller than in \Zn at high frequencies although the working frequency
used to investigate \Mg was 240~GHz while it was 34~GHz for \Zn. The distribution of the crystal field
does not affect this transition, and a distribution of g-factor should have as
effect to increase the linewidth when the frequency is increased.

\subsection{LT Phase}

\subsubsection{CW-EPR}

\begin{figure}[h]
	\centering
	\includegraphics{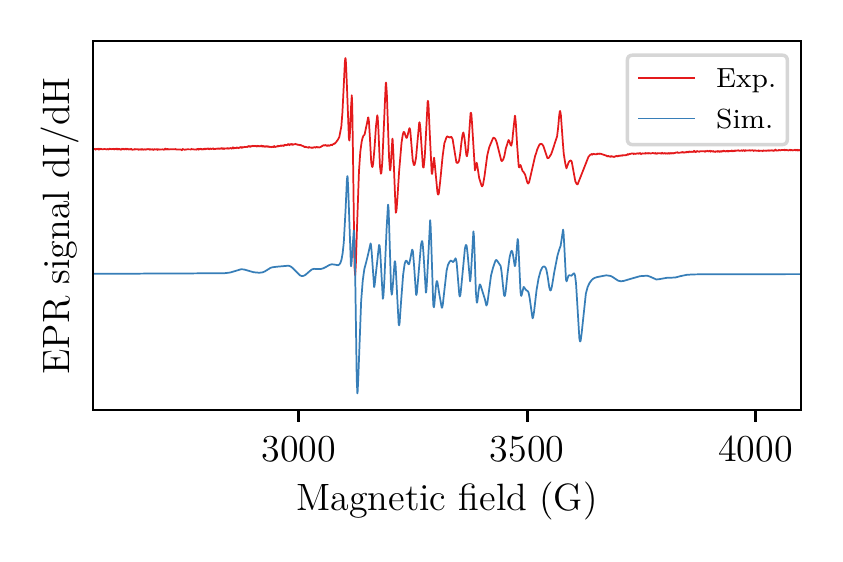}
	\caption{ First derivative if the EPR spectrum of \Mg recorded on a powder sample, at
		$T=100$~K and $f_{mw}=9.6$~GHz (red line). The best simulation obtained using
		the parameters reported in the main text is presented in blue. }
	\label{fig:eprpowderexpsim}
\end{figure}

When the temperature is below $T_c$=263~K, the system undergoes a structural
phase transition from disordered trigonal $R\bar{3}c$ to ordered monoclinic
$Cc$. Fig. \ref{fig:eprpowderexpsim} shows the EPR spectrum of \Mg powder at
T=100~K. The signal is now more resolved than in the HT phase due to an apparent
decreasing of the $\Delta D$ value.  Using  eq.\eqref{eq:Ham1} we simulate the spectrum
with the following parameters : $B_2^0$=110~MHz,  $B_2^2$=10~MHz, 
$B_4^0$=0.5~MHz, $B_4^2$=0.7~MHz, $B_4^4$=0.5~MHz and $A$=264~MHz. While second
order crystal field parameters are provided with a good approximation, the
fourth order ones have to be taken with caution since the transitions $m_s= \pm
5/2 \leftrightarrow  \pm 3/2$ and $m_s= \pm 3/2 \leftrightarrow  \pm 1/2$ are
not perfectly resolved. Nevertheless, the crystal field parameters are in the
range of what is expected for this family of material. \cite{Simenas2018}  The
crystal field distribution $\Delta D$ is about 50~MHz, much less than the one
needed to simulate the HT phase. In the LT phase, the DMA$^+$ cations now have only
one possible orientation induced by the ferroelastic phase. The residual $\Delta
D$ is thus due to local inhomogeneity.

\begin{figure}
	\centering
	\includegraphics{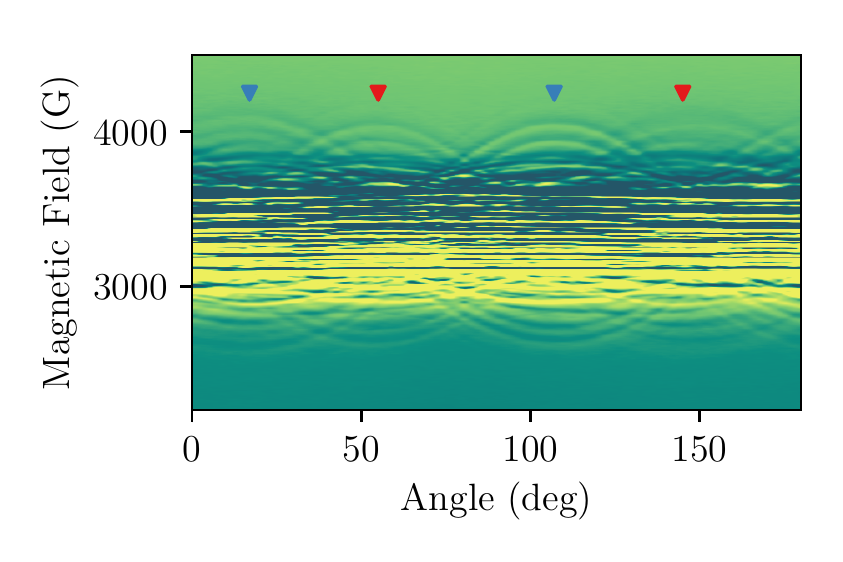}
	\caption{Angular dependence of the EPR signal of a single crystal of \Mg
		recorded at $T=100$~K and $f_{mw}=9.6$~GHz. The blue and red triangles show
		the
		extreme resonance field of the $m_s= \pm 5/2 \leftrightarrow  \pm 3/2$
		transitions (the only well resolved) and help to identify the two magnetic
		structures.  }
	\label{fig:eprxbandangle}
\end{figure}

Fig.\ref{fig:eprxbandangle} shows the angular dependence of the X-band EPR
signal of \Mg recorded at $T=$100~K with a resolution of 1$^\circ$. Most of the
transitions are between 3000~G and 3500~G and are therefor impossible to resolve. However, the highest anisotropic transitions, $m_s= \pm 5/2 \leftrightarrow  \pm 3/2$,
are visible on the edges of the spectra.  The triangles point to the maximum of resonance
fields and are found at 18$^\circ $ and 108$^\circ $ for the blue ones and at
55$^\circ $ and 145$^\circ $  for the red ones. The angular separation between
these two magnetic substructures is 53$^\circ$ which corresponds to the angle
between the two MnO$_6$ orientation resolved by XRD (see Fig. S1 \cite{Note1} ). It is worth noticing that,
in \Zn, six substructures was necessary to describe the
angular dependence of the EPR\cite{Simenas2017}. Single crystal XRD data obtained at 100~K reveals the presence of a two component twin with a minor domain of only 5\% weight. (Fig. S2 \cite{Note1}). 

\begin{figure}
	\centering
	\includegraphics{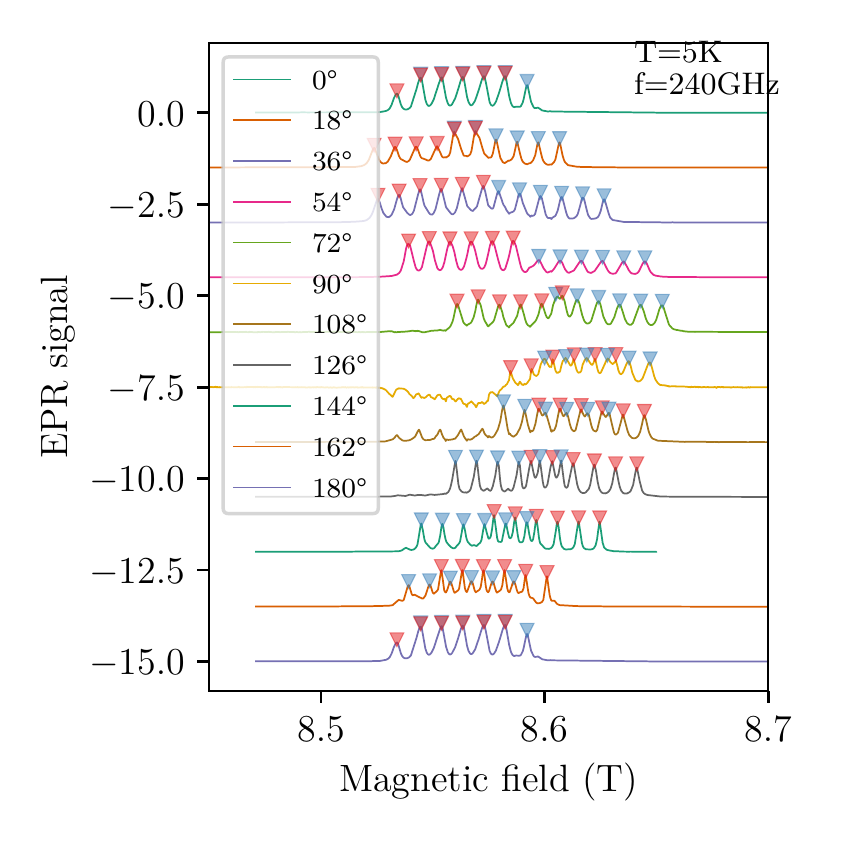}
	\includegraphics{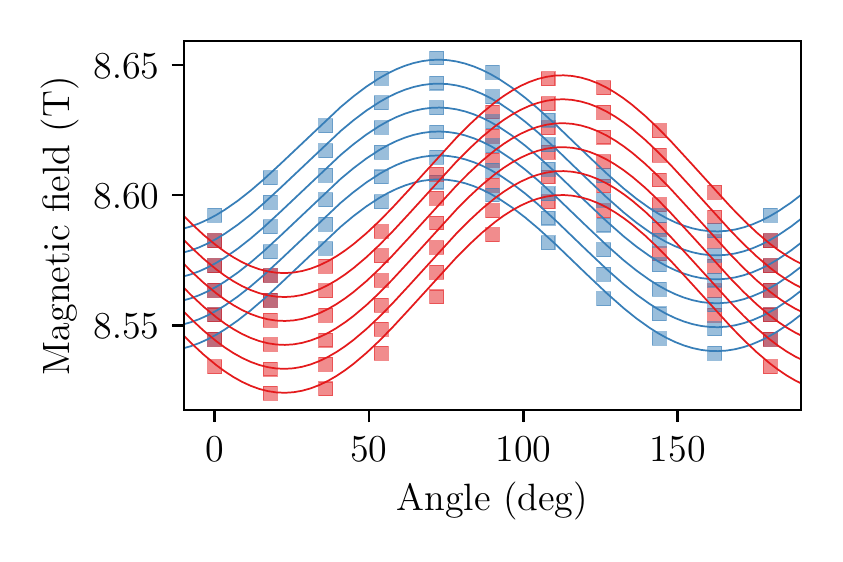}
	\caption{(top) Angular dependence of the HF-EPR spectra of a single crystal of
		\Mg recorded at $T=5$~K. Blue and red triangles indicates the resonance field.
		Temperature stability was not perfect and satellite lines from $m_s= -3/2
		\leftrightarrow -1/2$ can be seen at 90$^\circ$ and 108$^\circ$. (bottom)
		Angular dependence of the resonance fields. The squares represent the resonance
		fields
		and the lines correspond the simulations of the resonance field using Eq. \eqref{eq:Ham1} .  }
	\label{fig:serie240ghz}
\end{figure}

To confirm that \Mg has mostly a single ferroelastic domain, we performed high
frequency/field EPR (240~GHz $\equiv$ 8.87~T) at $T=$5~K. At this temperature and
for this field, the Boltzmann statistic populates mostly the lowest energy levels
($m_s=-5/2$). The transitions $m_s= -5/2 \leftrightarrow -3/2$ are mainly
visible, while the others are either weak or absent. Thus, the spectra are "cleaned"
and easier to read. Fig \ref{fig:serie240ghz} shows the angular dependence of
\Mg  at $f=240$~GHz.  Blue and red triangles represent the resonance fields and are useful to follow the angular variation of the resonance field of the two
magnetic substructures.   We can clearly distinguish the 2 sub-magnetic systems.
The small signals at 90$^\circ$ and 108$^\circ$ are due to thermal population of the 
$m_s=-3/2$ state which is caused by temperature instability in the cryostat.

While EPR studies \cite{Simenas2017b} has clearly shown the presence of three
crystallographic  twins of two magnetic domains in \Zn, we observe only a single
domain of two magnetic structures. If one consider that changing the metal ion in DMAXF from Zn to Mg modify the elastic property of the MOF, the same arguiment can be used to explain the dramatic  increase of $T_c$ in the DMAMgF compared to DMAZnF.

\subsubsection{Pulsed EPR}
\begin{figure}
\centering
	\includegraphics{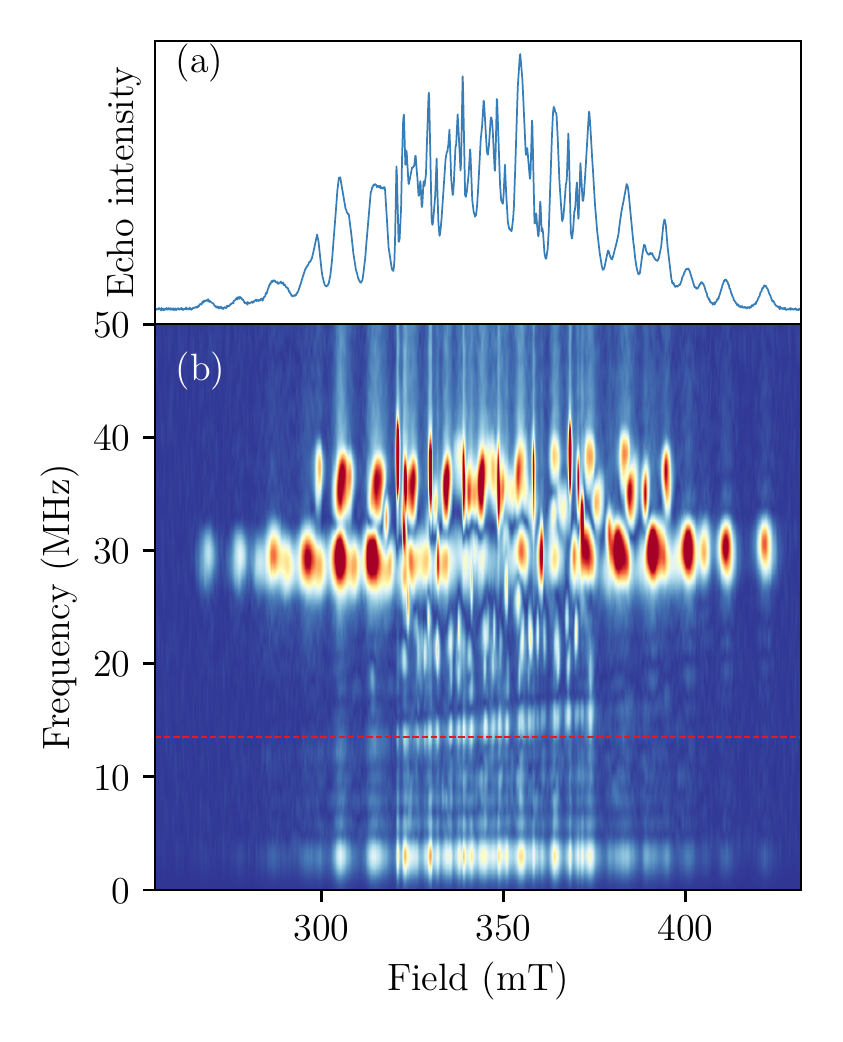}
	\caption{(a). Echo field sweep of a single crystal of \Mg  at $T=7$~K. (b) 
		Fast Fourier Transform of the field sweep Rabi oscillations for the same
		orientation with a pumping pulse of strength $h_{mw}=$4.8~G. Red (Blue) is the
		presence (absence) of frequency density. The dashed red line represents the
		frequency expected for $S=1/2$ at the same pulse amplitude.   }
	\label{fig:rabiexp2d}
\end{figure}

Despite our efforts to resolve all EPR lines in \Mg using low and high frequency
EPR, the large number of lines in Mn$^{2+}$ makes it difficult to identify
them completely. By means of pulsed EPR, we measured the field sweep Rabi
oscillations. The Rabi spectroscopy adds another dimension to the EPR spectrum.
While for a fixed field, the EPR intensity might contain many transitions with
unresolved contribution, the Rabi frequency of each transition  is often 
unique. On Fig. \ref{fig:rabiexp2d}(a) we show the 1D echo field and on Fig. \ref{fig:rabiexp2d}(b) we show the contour plot of the field sweep fast Fourier transform (FFT) of Rabi oscillations recorded on a single crystal of \Mg at $T=7$~K with
$h_{mw}=4.8$~G. The red dashed line corresponds to the frequency expected for a
$S=1/2$ spin. Clearly, the 2D Rabi field sweep helps to resolve many more transitions. The distribution
in the frequency dimension is due to the damping of the oscillation while the
distribution in the field dimension is due to the $\Delta D$. Qualitatively, the
broadly distributed frequencies are related to the $m_s= \pm 5/2 \leftrightarrow
\pm 3/2$ and $m_s= \pm 3/2 \leftrightarrow \pm 1/2$ transitions which are
sensitive to $\Delta D$ while the narrow distributed ones are related to  $m_s=
-1/2 \leftrightarrow 1/2$ which are insensitive to the crystal field. Moreover, in first
approximation, the Rabi frequency in a large spin system is given by the adapted
Fermi Golden rules \cite{Schweiger2001} : 
\begin{equation}\label{eq:rabi}
	F_R(S,m_s) = \sqrt{S(S+1)-m_s(m_s+1)} \times  F_R(S=1/2)  
\end{equation}
such as :  $F_R(S=5/2, m_s=-5/2)=\sqrt{5}\times  F_R(S=1/2)=30.2$~MHz, $F_R(S=5/2,
m_s=-3/2)=\sqrt{8}\times  F_R(S=1/2)=38.8$~MHz and $F_R(S=5/2, m_s=-1/2)=\sqrt{9}\times 
F_R(S=1/2)=40.5$~MHz

While the agreement with this model is fair (but not exact) for $m_s= \pm 5/2
\leftrightarrow \pm 3/2$ and $m_s= \pm 3/2 \leftrightarrow \pm 1/2$ it fails for
$m_s= -1/2 \leftrightarrow +1/2$. To understand the distribution of Rabi
frequencies in \Mg we developed a kernel density estimation of all transitions and
all orientations in the crystals. Using the crystal field parameters found in
the previous section, we computed the fields of resonance for each (allowed and
forbidden) transitions accessible from $H=$260~mT to $H=$430~mT. For each pair
of resonance field/transition we computed the amplitude of transition probability ($A_{tr}$) in the direction perpendicular to the field orientation that is
imposed by the experimental condition since the cavity force the microwave
polarization to be perpendicular to the static field. The Rabi frequency is then
$F_R=\frac{g\mu_b h_{mw}}{h}A_{tr}$, with $g$ is the g-factor (close to 2),
$\mu_b$ the Bohr magneton, $h$ the Planck constant and $h_{mw}$ the microwave
field. This series of Rabi frequencies is then computed for all orientations. We
discretize  the space using an icosphere to avoid over-density of orientations by
using a simple equally spaced Euler’s angles. The calculation included 20609
orientations and using a threshold of 10$^{-2}$ to suppress the far too low frequency
Rabi oscillations we obtained 1 346 898 sets of data containing resonance fields,
probability amplitude, transition identification and Euler angles. 

\begin{figure}[h]
	\centering
	\includegraphics[width=\linewidth]{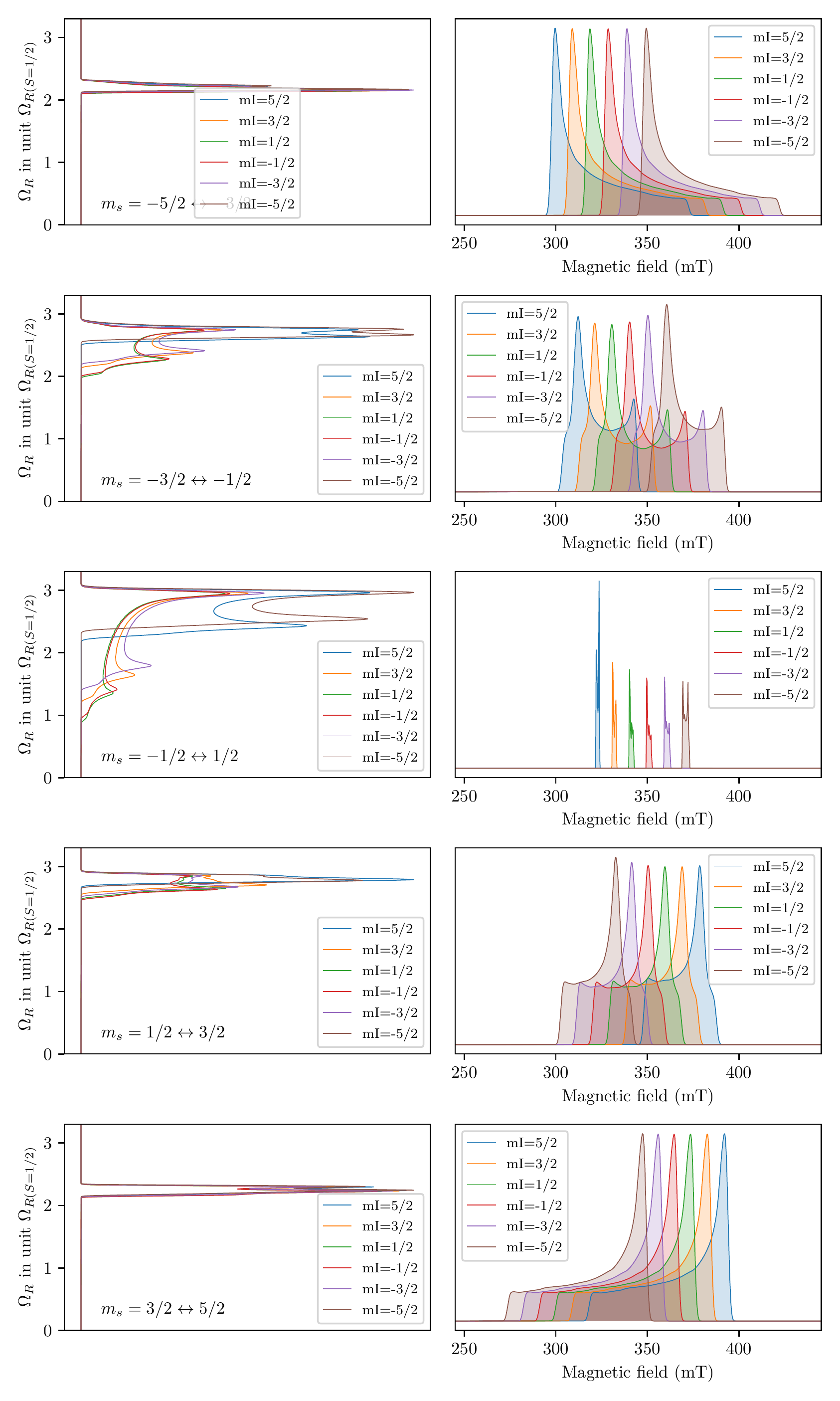}
	\caption{Rabi frequency (left) and resonance field (right) distributions
		calculated in \Mg using the crystal field parameters extracted from CW
		measurements for all allowed transitions ($\Delta m_s=\pm 1$ and $\Delta m_I=0$).
		The frequency distribution is presented vertically to help the comparison with
		the experimental data. The unit is set to  be proportional  to the Rabi frequency of
		the $S=1/2$ isotropic caliber. }
	\label{fig:rabidistcalcgrid}
\end{figure}

We first describe the Rabi and field distribution for the "allowed"
transitions ($\Delta m_s=\pm 1$ and $\Delta m_I=0$ ). Fig.
\ref{fig:rabidistcalcgrid} shows the Rabi frequencies and magnetic field
distributions for the 30 allowed transitions. First,  the field distribution
follows what we expected for a Mn$^{2+}$ ion in a moderate crystal field : $m_s=
\pm 5/2 \leftrightarrow \pm 3/2$ are spread on large distribution of the field,
$m_s= \pm 3/2 \leftrightarrow \pm 1/2$ are slightly less distributed while $m_s=
-1/2 \leftrightarrow +1/2$ are fairly insensitive to the crystal field and
induce the narrow lines observed on Fig. \ref{fig:rabiexp2d}. More surprisingly, the
Rabi frequency distribution is less intuitive. The transitions $m_s= \pm 5/2
\leftrightarrow \pm 3/2$ are weakly distributed around $2.2 \times F_R(S=1/2)$ whatever
$m_I$ is, confirming the validity of Eq. (\ref{eq:rabi}) while the $m_s= \pm 3/2 \leftrightarrow \pm
1/2$ transitions are different. $m_I=\pm 5/2$ also show a weak  frequency
distribution around $2.8 \times F_R(S=1/2)$ but $m_I=\pm3/2$ and $m_I=\pm1/2$ are much
more distributed. The transitions $m_s= -1/2 \leftrightarrow +1/2$ are even more
sensitive to the orientation. In this case the Rabi frequency distribution is very
broad and dependent to the  $m_I$ value while the resonance field distribution is
essentially independent to the orientation. This explains why the $m_s= -1/2
\leftrightarrow +1/2$ transition is narrow in the field dimension but
distributed in a large range of Rabi frequency (see Fig. \ref{fig:rabiexp2d}) .

Now we consider the case of Rabi frequencies and resonant field distributions
of the "forbidden" transitions. The calculation method is based on the first
order Fermi's golden rule and so only one photon is involved in the resonance
mechanism. The multiple photon transitions \cite{Sorokin1958, Bertaina2009,
	Bertaina2015a}  $\Delta m_s > 1$ are not taken into account. However such
transitions are induced only when $h_{mw}\sim D$ which is far from being the case
here. The "forbidden" transitions that we are considering are thus $\Delta
m_s=\pm 1$ and $\Delta m_I \neq 0$ . It appears that the transitions $ m_s=-1/2
\leftrightarrow 1/2 $ and  $ m_I=-1/2 \leftrightarrow 1/2 $  (see SI) are weakly
distributed in field, which is expected for such transition but we also observe that the
Rabi frequency distribution is centered at 1.6~$\times F_R(S=1/2)$ making these
transitions highly probable.

\begin{figure}
	\centering
	\includegraphics{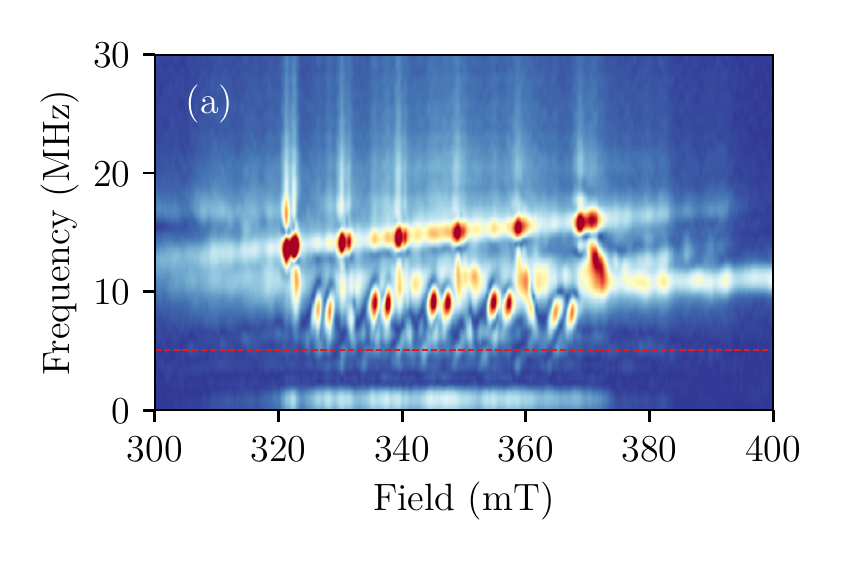}
	\includegraphics{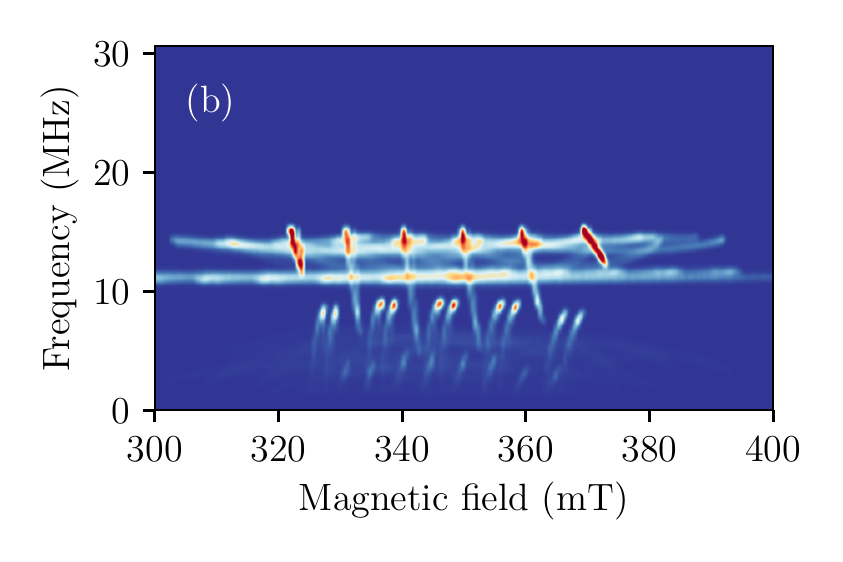}
	\caption{(a) Fast Fourier Transform of the field sweep Rabi oscillations for powder sample of \Mg. A pumping pulse of strength $h_{mw}=$1.8~G was used . Red (Blue) is the
		presence (absence) of frequency density. The dashed red line represents the
		frequency expected for $S=1/2$ at the same pulse amplitude. (b) Simulation of Rabi oscillations in a powder using the crystal field parameters of \Mg.  }
	\label{fig:rabiexp2dpowder}
\end{figure}

To confront the model, we have measured the field sweep Rabi oscillations of a
powder of \Mg. After some signal processing (baseline correction, hamming, zero
filing) the FFT is presented in Fig. \ref{fig:rabiexp2dpowder} (a). We then
simulated the distribution of both frequencies and resonance fields of a
Mn$^{2+}$ ion with the crystal field parameters extracted from CW-EPR of \Mg and
a microwave field of   $h_{mw}=$1.8~G. Our simulation describes rather well the
experimental data. We should note that the apparent slope in the Rabi
frequencies around 15~MHz in Fig. \ref{fig:rabiexp2dpowder}(a) is du to the
nuclear zeeman of protons $\omega_N$ (42.57~MHz/T ) which induce an
amplification of the Rabi intensity in case of Hartman-Hahn
conditions\cite{Hartmann1962} ($\omega_N \sim \Omega_R$ ). Without fitting
parameters we can describe the experimental data but the strength of this method
is to help of the identification of transitions. Fig. S6 and S7 show the
frequency and field distribution of all  principal transitions.

To simulate the single crystal field sweep Rabi frequency distribution presented
in Fig. \ref{fig:rabiexp2d} we just  replaced the Euler angles sweep by the
crystal field parameter D distribution in order to simulate the broadening. We
used a Gaussian distribution with width of 40~MHz corresponding to the strain of
$D$ extracted from EPR measurements.  We then calculate the field sweep Rabi
frequency distribution for two orientations separated by 58$^\circ$ which
correspond to the angle between the 2 magnetic domains observed in Fig.
\ref{fig:eprxbandangle} and presented in Fig.S8. The agreement between theory and experiment
(Fig. \ref{fig:rabiexp2d}) is fairly good thus confirming the presence of two
magnetic structures that are disoriented by about 60$^\circ$. We also note that very low
modes at about 2.5~MHz are visible experimentally but not  displayed in our model. We
believe these modes are related to the recently observed quantum rotor tunneling
of methyl group in DMAZnF\cite{Simenas2020} which has not been taken into account in our model.

\section{Conclusion}
We employed the electron spin resonance technique to investigate the dynamics of
the electron spin of a Mn$^{2+}$ ion used as a probe in the multiferroic
compound DMAMgF. In the high temperature phase, the X-band cw-EPR study
demonstrated that the correlation time describing the motion of DMA$^+$ in \Mg
is similar to that observed in the Zn analogue. The high frequency EPR data
revealed a large zfs distribution in \Mg. Using a wide range of configurations
and DFT computations, we were able to support this finding and fairly estimate
this distribution. In the low temperature phase, we observed a single elastic
phase containing two magnetic structures in agreement with single crystal XRD analysis. The
complex EPR structure observed in the LT phase was then solved using pulse EPR
combined to the field sweep Rabi oscillations method. A model based on the
crystal field and the kernel density estimation of all possible transitions and
orientations finally provided an accurate description of the complicated EPR
structure of \Mg.

\section*{Author's contributions}

SB and ND designed and directed the study. JKB synthetize the samples. JKB and JVT conducted the HF-EPR measurements. SB performed the X-band EPR measurements. MG performed the XRD measurements. MO carried out the DFT calculations. MO and SB conducted the theoretical analysis and wrote the paper with input from all authors. All authors contributed to the implementation of the research and to the analysis of the results.

\section*{Acknowledgments}
Pulsed ESR measurements  were supported by the Centre National de la Recherche Scientifique (CNRS) research infrastructure RENARD (Grant No. IR-RPE CNRS 3443). We thank the international research program of CNRS - PICS SomeTIME.  The NHMFL is supported by the NSF Cooperative Agreement Grant No. DMR-1157490 and the State of Florida.

\section*{Datasets}
The data that support the findings of this study are openly available in Zanodo at http://doi.org/10.5281/zenodo.4521882, reference number 4521882.

\section*{Supplementary information}
See supplementary information for XRD data and structure description, detail about DFT minimal models and simulation of rabi field sweep distributions.

\clearpage
%\bibliography{../FullBiblio.bib}% Produces the bibliography via BibTeX.
\bibliography{Main_preprint.bbl}
\end{document}